%% file: text.tex
\documentclass{elsarticle}
\usepackage{natbib}
\usepackage{color}
\usepackage{ulem}
\usepackage{graphicx}
\usepackage{rotating}
\usepackage{aas_macros}
\usepackage{subfig}
\usepackage{tabularx}
\usepackage[utf8]{inputenc}   

\newcommand{\nchan}{\ensuremath{\textnormal{N}_{\textnormal{chan}}}}
\newcommand{\npe}{\ensuremath{\textnormal{N}_{\textnormal{PE}}}}
\newcommand{\cxpe}{\ensuremath{\textnormal{CxPE}_{\textnormal{40}}}}

\begin{document}

\title{Sensitivity of the High Altitude Water Cherenkov Detector to Sources of Multi-TeV Gamma Rays}

\input{hawc-authors-elsevier-initials}

\begin{abstract}

The High Altitude Water Cherenkov (HAWC) observatory is an array
of large water Cherenkov
detectors sensitive to gamma rays and hadronic cosmic rays
in the energy band between 100 GeV and
100 TeV.
The observatory will be used to measure 
high-energy protons and cosmic rays via detection of 
the energetic secondary particles
reaching the ground when one of these particles
interacts in the atmosphere above the detector.
HAWC is under construction at a site 4100 meters above sea level on the
northern slope of the volcano Sierra Negra, which is located in central
Mexico at 19$^{\circ}$N latitude.
It is scheduled for completion
in 2014.
In this paper we estimate the sensitivity of the HAWC instrument
to point-like and extended sources of gamma rays. The source
fluxes are modeled using both unbroken power laws and power laws with
exponential cutoffs. 
HAWC, in one year, is sensitive to 
point sources with
integral power-law spectra
as low as $5~\times~10^{-13}$~$\rm{cm}^{-2}~\rm{sec}^{-1}$ 
above 2 TeV
(approximately 50 mCrab)
over
5 sr of the sky.
This is a conservative estimate based on simple event parameters
and is expected to improve as the data analysis techniques are refined.
We discuss known TeV sources
and the scientific contributions that HAWC can make to our
understanding of particle acceleration in these sources.

\end{abstract}

\maketitle
\section{Introduction}

TeV astronomy began in 1989 when gamma rays from the Crab Nebula were
recorded at the Whipple Observatory during 81 hours of
observation \cite{whipplecrabdiscovery}.  Due
to the very low flux of TeV gamma rays at Earth (about
$10^{-11}$ photons cm$^{-2}$ s$^{-1}$ TeV$^{-1}$ at 1~TeV)
the Whipple
telescope observed gamma rays indirectly by imaging the Cherenkov light
produced in the atmosphere by gamma-ray extensive air showers.  Since 1989,
the number of confirmed TeV sources has grown to about 140
objects \cite{tevcat},
with most of the observations made by Imaging Atmospheric Cherenkov Telescopes
(IACTs) based on the Whipple design.  The current generation of IACTs -
HESS \cite{hesscrab},
VERITAS \cite{veritascrab}, and MAGIC \cite{magiccrab} --
are sensitive enough to measure the
Crab Nebula at the 5$\sigma$ level with only several minutes of
observations.

A second method of observing TeV gamma rays, known as the water Cherenkov
technique, was developed at the Milagro Observatory \cite{milagrocrabspectrum},
which operated from
2000 to 2008 in the mountains above Los Alamos, NM.  Rather than image the
Cherenkov light from extensive air showers, the Milagro detector was
designed to directly sample shower particles at ground level using a large
pool of water in an optically isolated reservoir.  The reservoir was 
instrumented
with an array
of photomultiplier tubes (PMTs) to record the Cherenkov light produced by
charged particles entering the water.

For each air shower observed in the array, the
timing and amplitude of signals in the PMTs can be used to estimate the
arrival direction of the primary particle and the position of the shower
core.  With this technique it is also possible to discriminate showers
produced by photons from air showers produced by cosmic-ray hadrons, because 
hadronic showers characteristically 
produce isolated regions of high energy deposition far from
the shower core.
While the Milagro detector was less sensitive to point sources than the
current generation of IACTs, it had the advantages of $>90\%$ uptime and
a large instantaneous field of view.  As a result, the experiment was 
used to
discover new Galactic sources of TeV gamma rays \cite{milagrogalacticplane},
diffuse emission
from the Galactic plane \cite{milagrodiffuse} \cite{milagrocygnus},
and measure an anisotropy in the arrival
direction distribution of charged cosmic rays
\cite{milagrointermediate} \cite{milagrolsa}.

Following the success of Milagro,
the High Altitude Water Cherenkov detector (HAWC) is a water Cherenkov
extensive air shower array under construction at Sierra Negra, Mexico.
HAWC will consist of an array of 300 water Cherenkov detectors (WCDs).
Each WCD comprises a steel tank 7.3 m in diameter and 4.5 m in height, a
plastic bladder to contain 200~000 liters of purified water, and four PMTs:
three 8-inch Hamamatsu R5912 PMTs re-used from Milagro and one 10-inch R7081-MOD
high-quantum efficiency PMT.  The WCDs will be deployed in a close-packed
array over an area of approximately 20~000 m$^2$.
The PMTs detect Cherenkov
light from energetic particles passing through the WCDs. 
The air shower core is determined by a fit to the amount of light
detected by each PMT and the direction of the incident particle is fit
by the time (measured to $\sim$2 ns) each PMT is hit.

The HAWC instrument, the reconstruction and analysis algorithms,
and the sensitivity to
gamma-ray bursts, focusing on emission below 1 TeV,
have been described previously \citep{hawcgrbsensi}.
In the following sections we present
the results of a study of the sensitivity of HAWC to
steady sources of gamma rays from 1 TeV to 100 TeV.

\section{Analysis Technique and Cuts}

When observing gamma-ray sources in the 100 GeV - 100 TeV range,
the dominant background comes from the
abundant population of hadronic cosmic rays,
mostly protons and helium nuclei.
These cosmic rays arrive
nearly isotropically,
with anisotropy
present at a $10^{-3}$ level
\citep{milagrointermediate}\citep{milagrolsa}.
A source of gamma rays appears as a small localized bump
on top of this smooth background.
The analysis of a potential
source consists of defining a small angular bin around the position of the
source, determining the number and statistical uncertainty of background
events that will appear in this angular bin, and measuring the
number of events above background in the bin.
We attempt to remove hadronic events with a set of cuts intended to
identify the penetrating particles (mostly muons) characteristic of
a hadronic air shower.

Simulations are used to model the air shower
and ground detector components. 
The air shower is modeled using CORSIKA \citep{corsika} and the
detector components are modeled with Geant4 \citep{geant4}.
The reconstruction and gamma/hadron
separation algorithms are applied to the simulated output.
The calculation of sensitivity to high-energy steady sources is
similar to the analysis described in  \cite{hawcgrbsensi}.
Since the publication of a previous study\cite{hawcgrbsensi}, 
the design of HAWC has been modified to include
an additional 10-inch
high-efficiency PMT at the center of each WCD. The current
sensitivity computation includes this PMT modeled
as one of the re-used Milagro 8-inch PMTs. This 
provides a conservative estimate
to the sensitivity while deferring the issue
of developing appropriate algorithms to handle the two distinct
types of PMTs.

We use
the core and angular
reconstruction algorithms and calculate a ``compactness''
parameter for photon/hadron discrimination as in \cite{hawcgrbsensi}.
The compactness parameter is designed to identify muons
in the air shower, which appear as localized
charge depositions.
The compactness is
defined as $\nchan{}/\cxpe{}$ where \nchan{}
is the number of PMTs participating in an event and
\cxpe{} is the total number of photo-electrons (PEs)
in the PMT with the largest signal that is located
outside a radius
of 40 meters from the reconstructed air shower core.
However, the
optimization of the cuts for the present sensitivity calculation
differs from \cite{hawcgrbsensi}
because we are primarily interested in high-energy
sources with spectra extending beyond 1 TeV.

The performance of the HAWC detector,
notably the angular resolution and gamma/hadron separation,
improves
with the size of the air shower on the ground.
Larger air showers are
better measured and easier to reconstruct. Furthermore, the
number of muons in a cosmic ray air shower increases with the energy
of the incident cosmic ray, and so the gamma/hadron
discrimination also improves with energy.
After hadron removal cuts and a cut
on the angular distance of events to a source, we
anticipate (for a Crab-like point source source) a signal-to-background
ratio of between 1:350 at 100 GeV to about
10:1, at 10 TeV. 
The event parameters used here are
relatively simple. We anticipate
the sensitivity to improve beyond what is presented here as
the event reconstruction is
improved. 

In order to account for the energy dependence of the
sensitivity, the data are divided into 14 bins using
\nchan{}, the total number of PMTs hit during an event, and
\npe{}, the total number of photo-electrons (PEs) measured
by the PMTs during the event. This binning is somewhat arbitrary
but its selection does not impact the sensitivity appreciably.
Both \nchan{} and \npe{} are correlated with shower
energy, but neither is an ideal energy estimator because they
do not account for the atmospheric slant path of the shower geometry,
nor do they account for how well-contained ths shower is within the array.
More sophisticated energy estimation
algorithms are expected in the future.
The bins are arranged so that
they reflect larger -- and thus higher-energy --
events.

The 14 bins chosen for \nchan{} and \npe{} are shown
in Table \ref{binstable}.
Bins 0-8 are determined using \nchan{}. Higher bins are
determined by \npe{} because \nchan{} saturates
as an energy estimator when $\sim$3/4 of the array is hit
For each bin, we determine the 
angular bin and compactness cut that maximizes the statistical significance
of that bin.
The optimization is made by maximizing the
statistical significance of a hypothetical source with
a differential photon flux\cite{hesscrab} of
\begin{equation}
{dN/dE}=3.45 \times 10^{-11} (E/{\rm TeV})^{-2.63}~{\rm cm^{-2}~s^{-1}~TeV^{-1}}
\label{hesscrabflux}
\end{equation}
and a declination of +35$^{\circ}$. The declination +35$^{\circ}$ is chosen
because it is characteristic of sources in the field-of-view 
of HAWC, which is located at 19$^{\circ}$N. The spectral index
$\alpha=-2.63$ has been chosen as a reasonable compromise between 
sources with photon fluxes concentrated at the low and high-energy
limits of the HAWC energy range.

\begin{sidewaystable}
\begin{tabular}{c|c|c||c|c||c|c|c|c|c}
     &            &                        & Angular & Compact.    &          &         &                &                    \\
     & $\nchan{}$ & $\rm{log_{10}(\npe{})}$ & Bin     & Cut         & Signal   & Weight  & $E_{\rm{log}}$  &  \\
 Bin & Bin        & Bin                    &  (deg.) &  (PE$^{-1}$) &  $S_{i}$  & $w_{i}$ & (GeV)  &        $\sigma_{E_{\rm{log}}}$                \\
\hline
\input{bins}
\end{tabular}
\caption{
The table shows the \nchan{} and ${\rm log}_{10}{(\npe{})}$ values
used to define the analysis bins
along with
angular bin and compactness cut chosen
to optimize the statistical significance 
of the hypothetical source in Equation \ref{hesscrabflux}
at +35$^\circ$ declination.
Also shown are
the number of photon events from our hypothetical source
passing cuts,
and the
weights used in Equation \ref{sigmacalc} optimized for 
the hypothetical photon flux given in 
Equation \ref{hesscrabflux}.
Also shown are the median and width, in logarithmic
space,
of the true photon energy
distribution of events
$E_{{\rm log}} = {\rm log}_{10}(E/{\rm GeV})$ in each bin.
The results are presented for one year of data taking.}
\label{binstable}
\end{sidewaystable}

To calculate the significance of a point
source, we assume that we are measuring
the source at a fixed declination over many transits of the source.
We weight
the simulated events by the amount of time spent at each zenith in order
to account for the source transit. We then calculate the number of
expected signal and background events in our analysis bins.
Table \ref{binstable} shows the results of this
optimization.

For a given hypothetical source, we wish to compute the significance of
one year of observations while accounting for the changing signal and
background efficiencies in each bin.  To do so, we assign a weight
$w_i=S_i^{\rm opt}/B_i^{\rm opt}$ to each analysis bin $i$, where
$S_i^{\rm opt}$ and $B_i^{\rm opt}$ are the expected signal and background
from
the optimization hypothesis.
If the expected
signal and background counts from a given source are $S_i$ and $B_i$, which
could in general be different than for the $S_i^{\rm opt}$ and $B_i^{\rm opt}$
of the optimization hypothesis, then
we calculate the significance of the observation as
\begin{equation}
\sigma = \frac{\sum_i w_i\cdot S_i}{\sqrt{\sum_i w_i^2 B_i}}
\label{sigmacalc}
\end{equation}
where the sum is taken over the 14 analysis bins.

\begin{figure}
  \includegraphics[width=60mm]{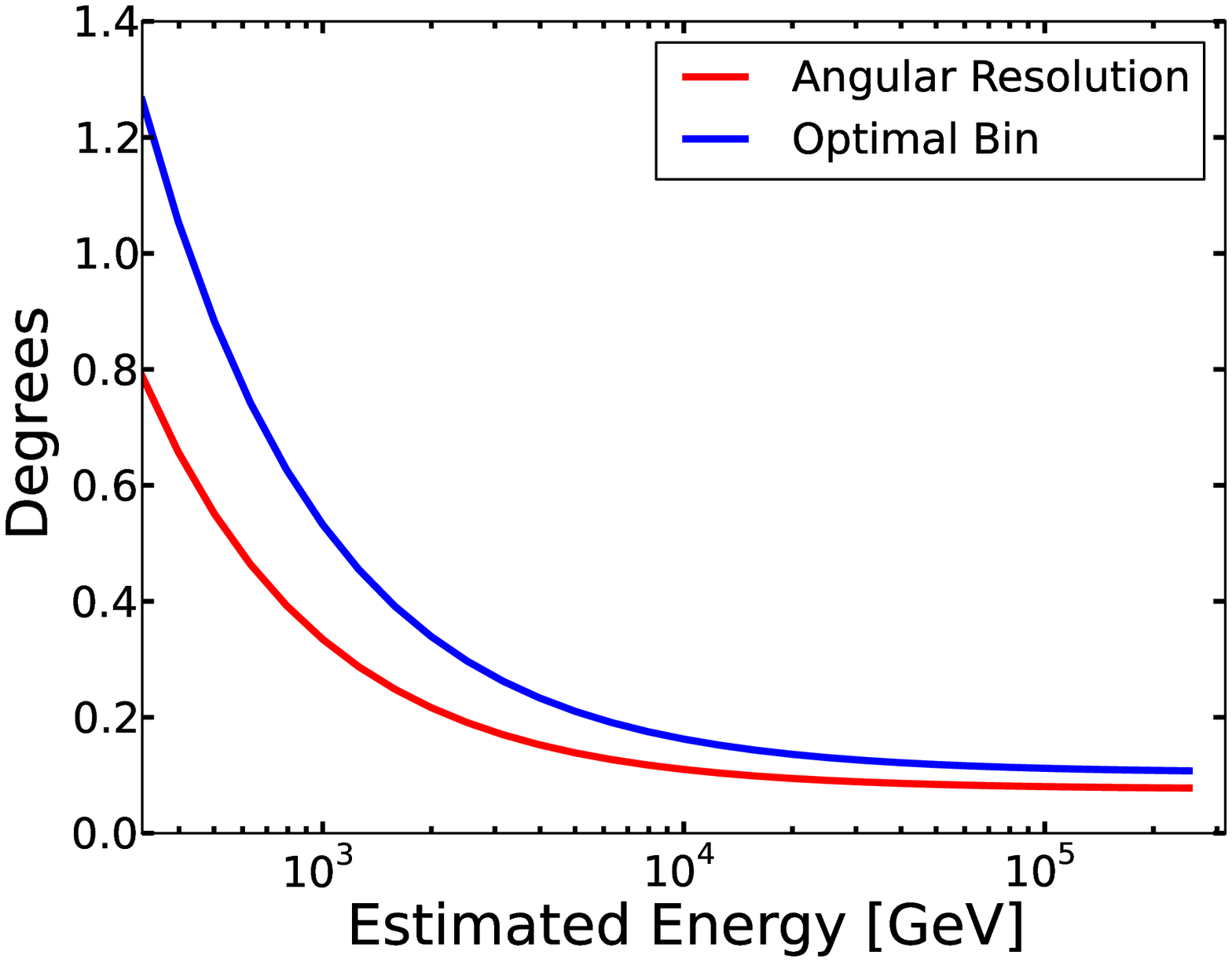}
  \includegraphics[width=60mm]{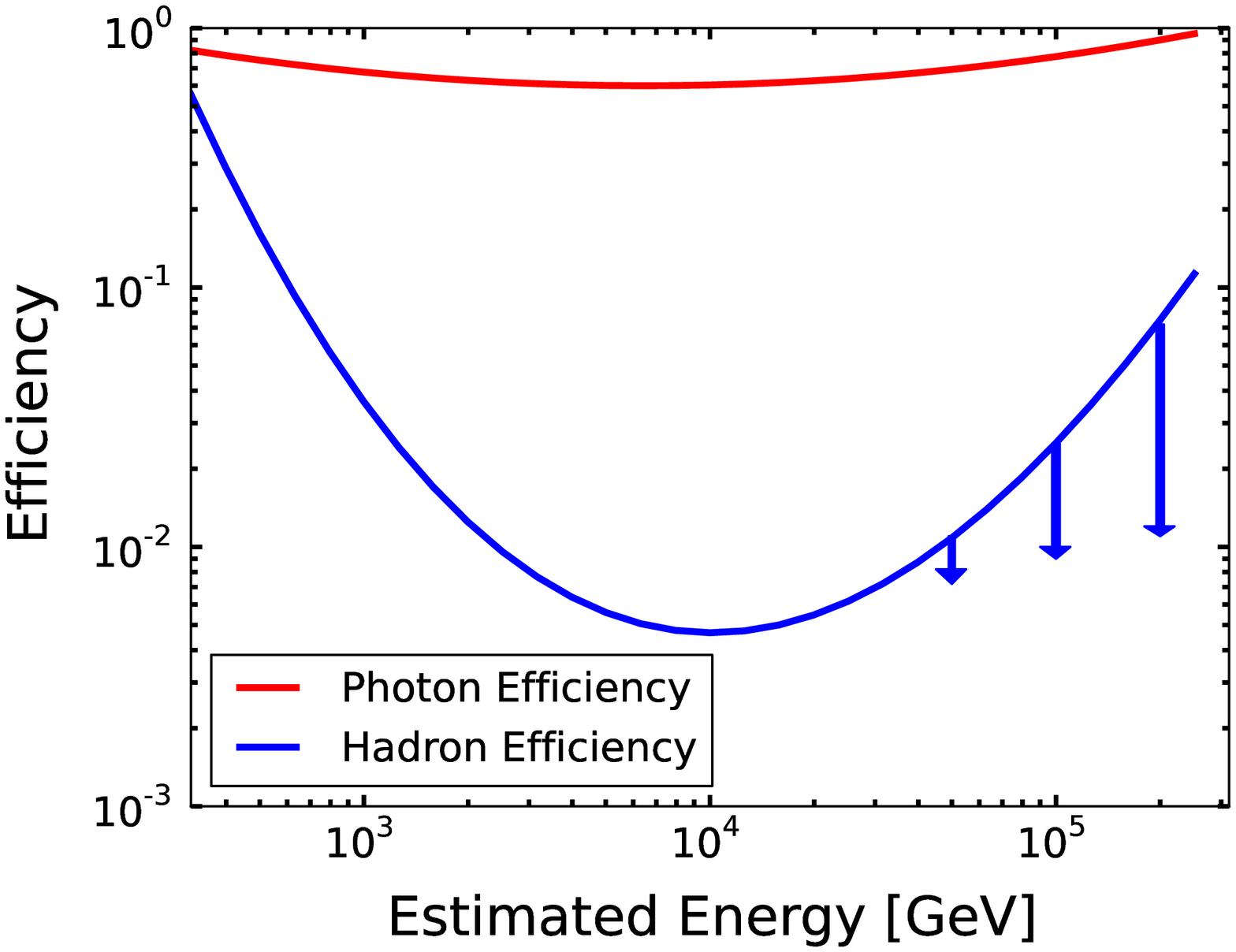}
  \caption{Performance of the optimal cuts from Table \ref{binstable}.
The left figure shows the angular resolution (one standard deviation
of a fitted two-dimensional Gaussian)
and optimal angular bin (column 4)
for the energy bins from Table \ref{binstable}, smoothed for display.
The figure on the right shows the efficiency of the compactness
cut, after all other cuts have been made, on the photon signal
and background in the bins from Table \ref{binstable}.
Beyond $\sim$10 TeV the photon discrimination becomes strong enough
to remove all of the existing simulated background events, and we
conservatively
require 10 simulated events to survive cuts so the background can be reliably
estimated.
The arrows above 10 TeV emphasize that the hadron rejection is
anticipated
to improve.
}
  \label{performance}
\end{figure}

Figure \ref{performance} shows the performance 
of the angular localization and photon discrimination 
after applying the optimized cuts from Table \ref{binstable}. 
As one can see, the angular resolution improves steadily with increasing
energy, reaching $\sim$0.1$^\circ$ at E~$>$~10 TeV. 
The hadron rejection efficiency improves with increasing energy up to about
10 TeV, at which point none of the simulate background survive cuts. This
makes the analysis difficult to optimize at high energy. As a conservative
approach, we choose the cut that preserves at least
10 simulated background events to estimate the hadron rejection efficiency. 
This approach is likely to produce an underestimate in the predicted 
sensitivity at high energies. 
The issue will be solved when the experiment starts taking data because the
data (dominated by hadronic background) can be used to estimate the 
rejection directly.
Nevertheless the HAWC sensitivity above 10 TeV 
is very close to the
limit
implied by the need to detect at least $\sim$10 events.

Since many Galactic sources are extended,
we also consider the sensitivity of the instrument to sources that are
extended by some amount. We presume that the spatial extent
of the source is a disk with a radius
$R_{S}$ at all photon energies.
The measured distribution is then the convolution of the true source
distribution
and the response of the instrument to a point source.
Once the source is smeared by the point response of the instrument,
the optimal angular bin
is calculated to maximize Equation \ref{sigmacalc}. In general,
the optimal bin size of an extended source will be larger than that of
a point
source, meaning more background is admitted. All other cuts remain the same.

\section{Results}

In our analyis, spectra of the form 
\begin{equation}
\rm{dN/dE = \Phi_{0} (E/TeV)^{-\alpha} exp(-E/E_{cut})}
\end{equation}
are considered, where $\rm{\Phi_0}$ is the flux at 1 TeV, $\alpha$ is the
source spectral index and $\rm{E_{cut}}$ is the energy at which the spectrum
cuts off. 
While a source may have a more complicated energy spectrum, the
power-law with an exponential cutoff has been sufficient to describe most
TeV sources to date.

\begin{figure}
\includegraphics[width=110mm]{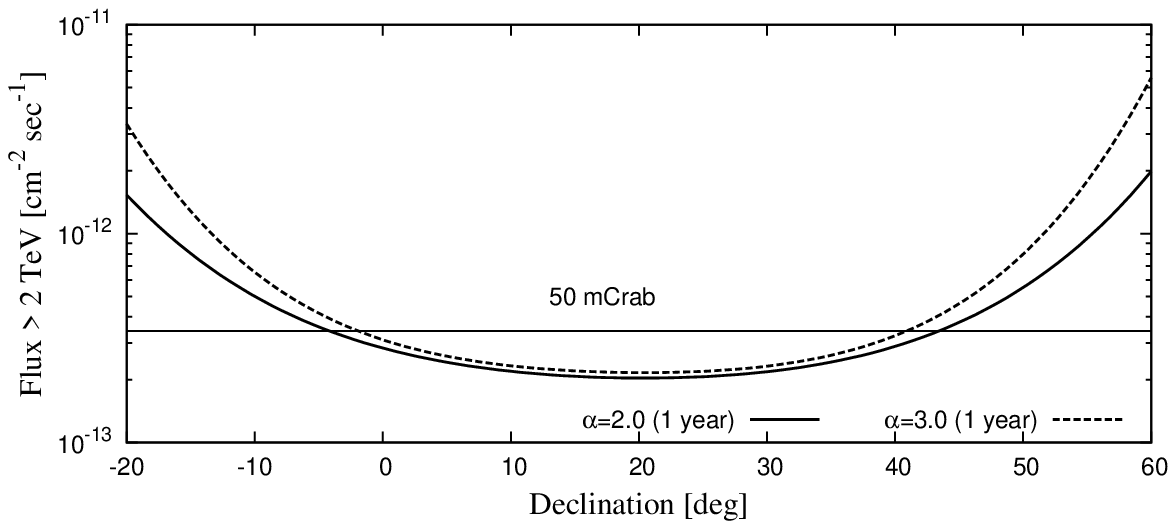}
\includegraphics[width=110mm]{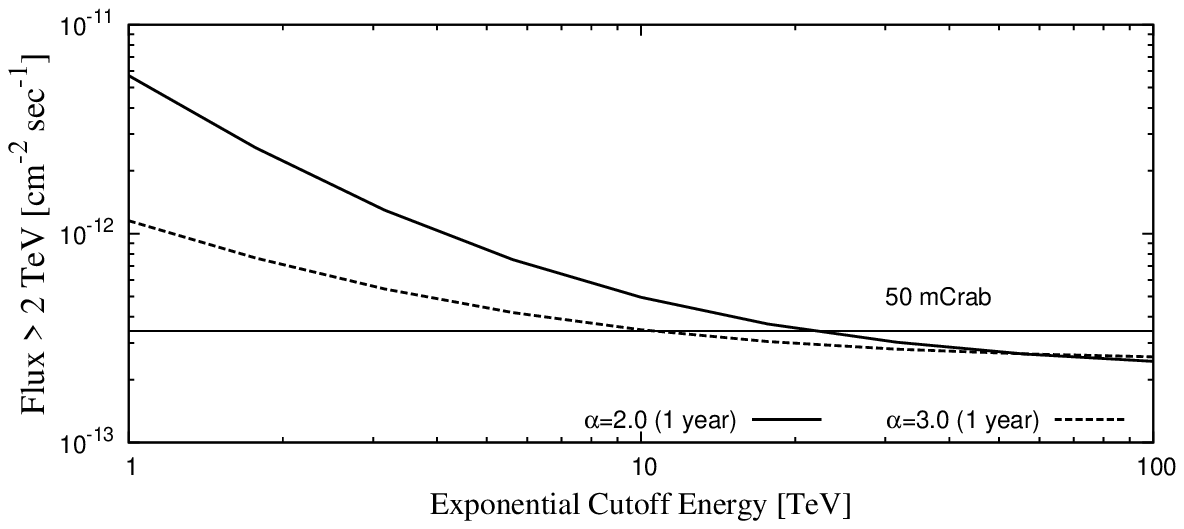}
\includegraphics[width=110mm]{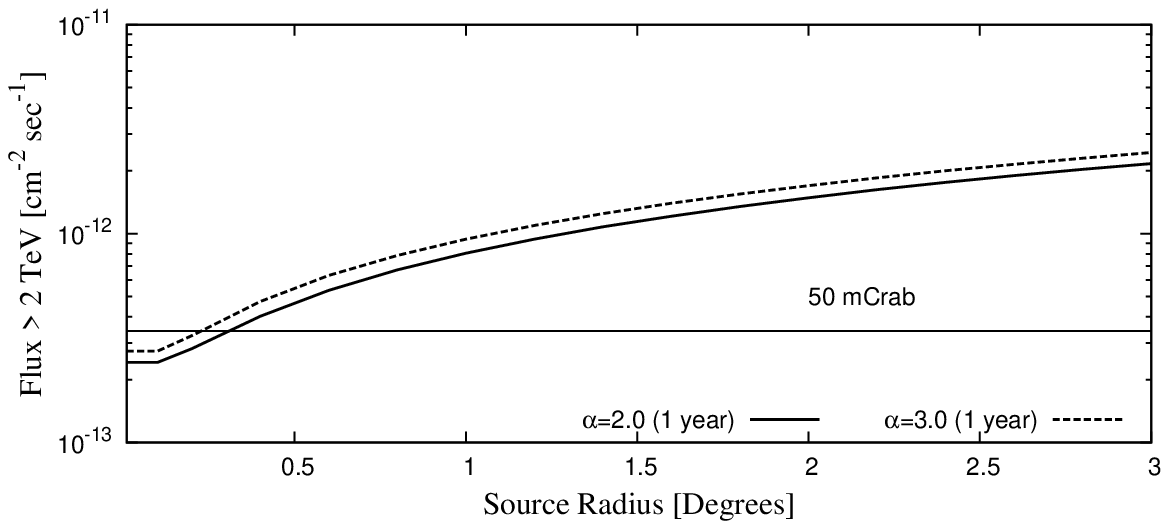}

    \caption{The sensitivity of HAWC to sources with varying spectral
      parameter. We show the flux required to give a central
      expectation
of 5$\sigma$ in one year.
 The top panel shows the sensitivity of HAWC to
      sources with pure power-law spectra as a function of declination.
      In the center panel, the sensitivity to a source at $+35^{\circ}$
      declination is shown as a function of spectral cutoff energy.  In the
      bottom panel, the sensitivity to a source (also at $+35^{\circ}$) with
      a pure power law spectrum is shown as a function of the spatial extent of
      the source.  Note that one day of data corresponds to one transit of the
      source, which means the source spends only a few hours in the field of
    view of the detector. For the computation of the integral Crab flux,
we have assumed the pure power-law measurement from \cite{hesscrab} 
which is the same as is given in Equation \ref{hesscrabflux}.
} \label{sensi-point}

\end{figure}

Figure \ref{sensi-point} shows the sensitivity of HAWC to sources
of varying spectral and spatial parameters.
To be considered within the sensitive range of the instrument the mean signal
expectation must
be detectable at 5 standard deviations (Equation \ref{sigmacalc})
above background with a full year of data. Note that one year of data
presumes one year's worth of transits of the source overhead and 
includes a substantial amount of time with the hypothetical source
out of the instrument's field-of-view.
The top panel of Figure \ref{sensi-point} shows
the sensitivity of the detector to sources with no
cutoff ($E_{\rm cut}=\infty$)
as a function of the source's declination.
The center panel shows the sensitivity to a source at +35$^{\circ}$ declination
as a function of cutoff energy assuming two different values for the
differential spectral index.
The bottom panel shows the sensitivity to a
source at +35$^{\circ}$ declination as a function of the spatial extent of the
source 
assuming a pure power law spectrum with no cutoff.
The flux sensitivity is expressed as the integral flux above
2 TeV. This choice nearly 
eliminates the dependence of the sensitivity of pure
power-law spectra on the spectral index.

\begin{figure}
\includegraphics[width=110mm]{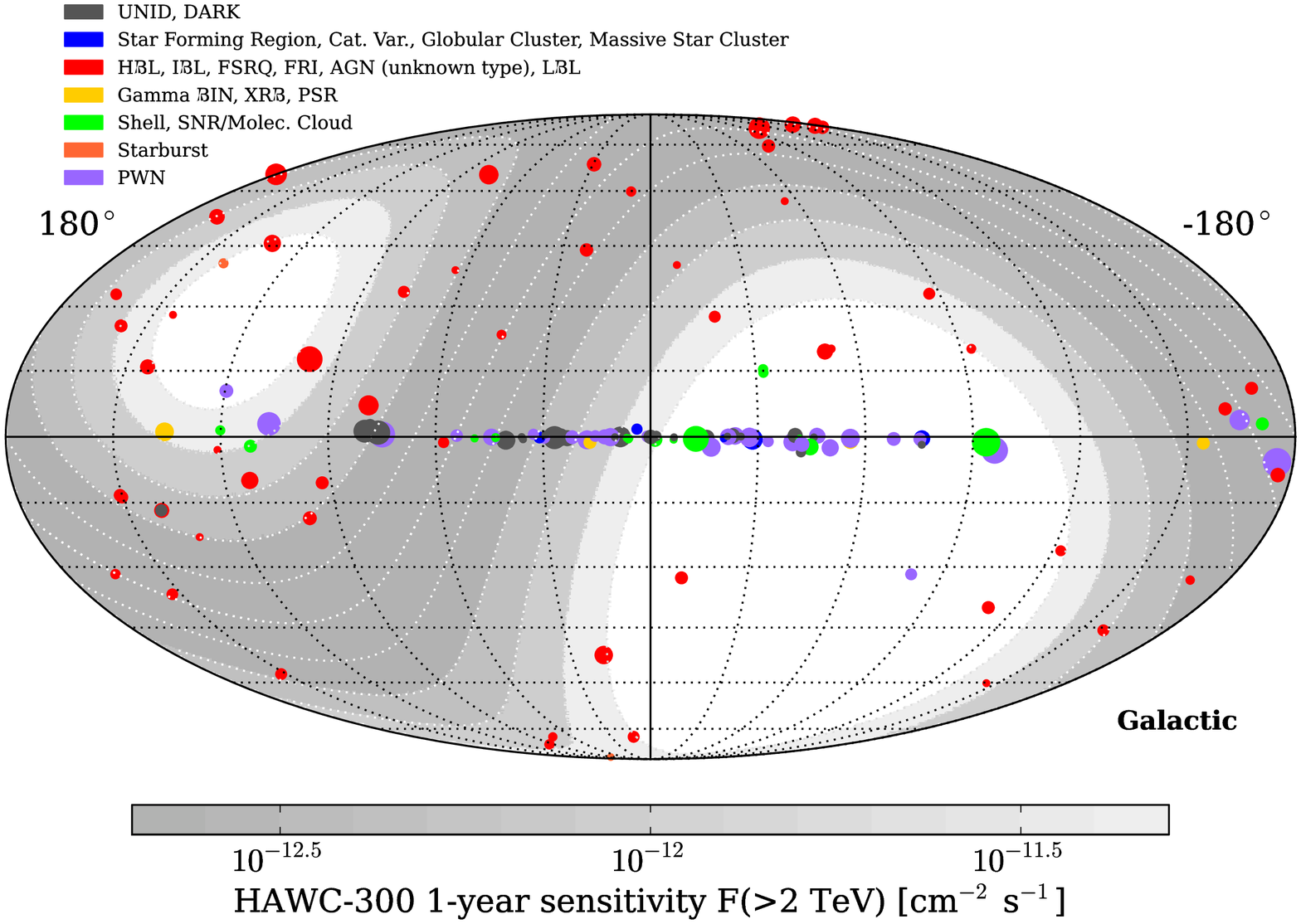}
\caption{The sensitivity of HAWC to differential 
$E^{-2}$ spectra along with known
  sources
from the TeVCat TeV catalog \cite{tevcat} in Galactic coordinates. The 
size of the source is representative of the source's flux in the TeVCat 
catalog.}
\label{sensi-galactic}
\end{figure}

We find that over one year of exposure, 
HAWC is sensitive to pure power law spectra
at a level of 
 $5 \times 10^{-13}$ $\rm{cm}^{-2} \rm{sec}^{-1}$ ($\sim$ 50 mCrab) 
above 2 TeV
 over 5 sr (40\%) of the sky. Figure
\ref{sensi-galactic}
shows the sensitivity to differential $E^{-2}$ spectra in Galactic coordinates
along with sources from the TeVCat catalog \cite{tevcat}.
It is worth noting that the form of Equation \ref{sigmacalc}
indicates that if a source were to flare by a factor of $N$,
the time
required to see the source at fixed significance would be reduced by $N^2$.
Similarly, since HAWC is roughly 15 times more sensitive than Milagro, we
anticipate to see the Milagro sky $\sim$225 times faster, i.e. 
in just over a week of data-taking with
the full HAWC instrument.

\begin{figure}
    \includegraphics[width=120mm]{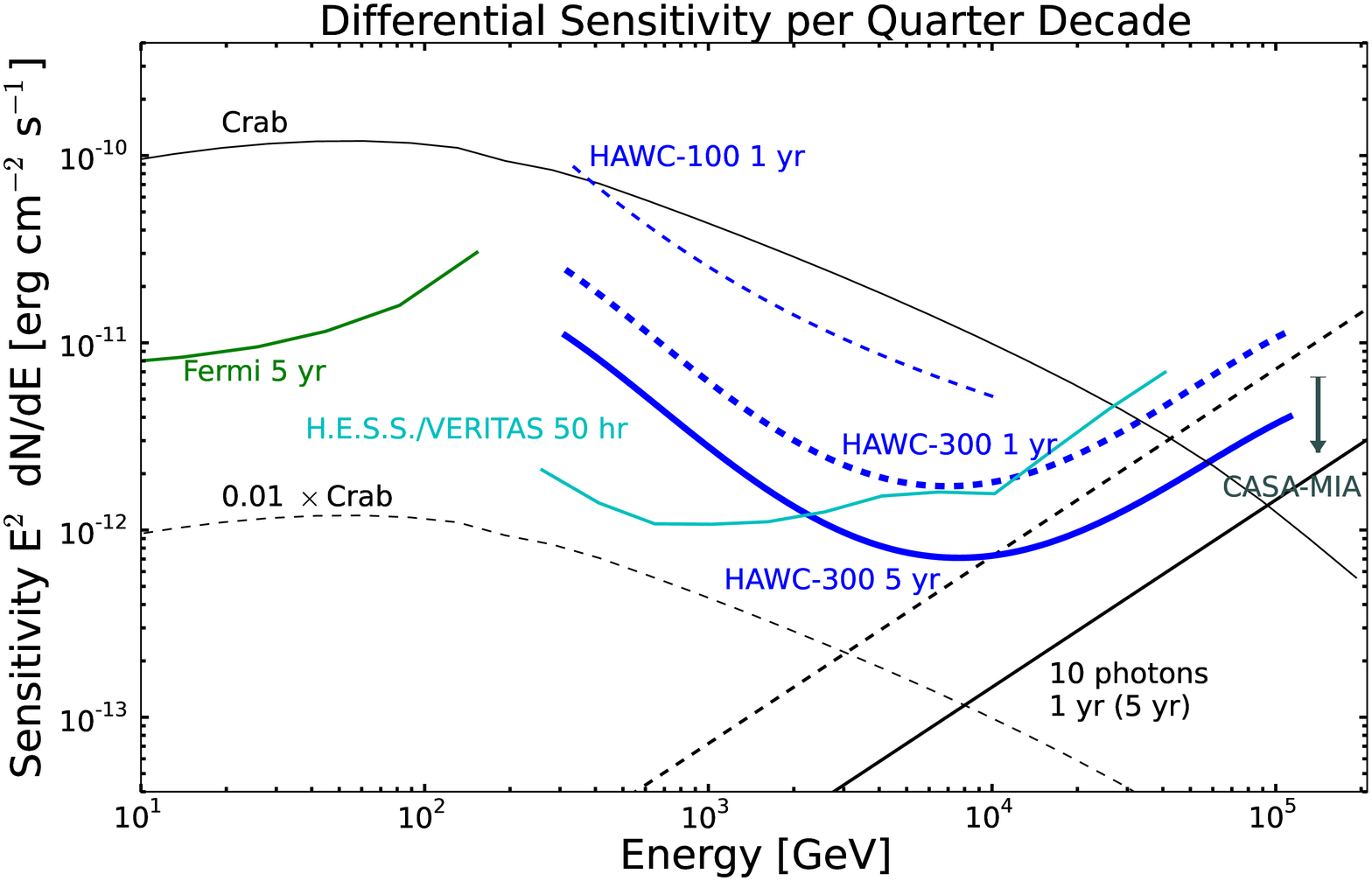}
  \caption{The differential sensitivity of HAWC compared to peer
  instruments. The sensitivity is found by calculating the flux needed for
  a 5$\sigma$ detection for quarter-decade energy bins using Eqn. 17 from
  \cite{lima}. Quarter-decade energy bins are found by normalizing the bins in
  Table \ref{binstable} to facilitate comparison to 50-hour
IACT\cite{hesscrab}\cite{magicperformance}\cite{veritasperformance}
 and 3-year Fermi\cite{fermiperformance} sensitivity
 calculations. Note that the actual triggering threshold
extends somewhat lower than implied by this figure, down to 100 GeV.
We show the flux required to produce 10 events in 20~000 m$^{2}$
assuming a 5 hour observation each day and 40\% loss of efficiency due 
to angular bin cuts and a 50\% loss of efficiency due to photon/hadron
discrimination.
Without improvement to the high-energy effective
area,
this is a practical limit to the HAWC sensitivity.
  Also shown, for reference, 
is the measured Crab flux from \cite{hegracrab} and 
an inferred quarter-decade differential upper limit from the Crab at 
141 TeV from the
CASA-MIA experiment \citep{casamiacrab}. The sensitivity for the first 100 of
the 300 HAWC WCDs is shown for comparison.
}
  \label{sensi-diff}
\end{figure}

The HAWC differential sensitivity is shown alongside the sensitivity from
other current
instruments in Figure
\ref{sensi-diff}.
HAWC complements these instruments with its high-energy reach
and its wide field-of-view. Whereas an IACT must point at a given source on
dark, clear nights
in order to observe it and can observe $\sim$15 sources with
50 hours of exposure in a year, 
HAWC continuously records events across the entire
overhead
sky. This capability is useful for the recording of transient
source events,
and assists in the energy reach of the instrument above 1 TeV.
The power-law
spectrum of a typical astrophysical source means extreme high-energy events
are very rare,
making continuous observations important.

\begin{sidewaystable}
  \begin{tabular}{c|c|c|c|c|c|c|c|c}
 Name & RA & Dec & $\alpha$ & $E_{cut}$  & Flux ($>$ 2 TeV)& Significance & Reference \\
      & (deg)&(deg)&          &    (TeV)  & 1$\rm{x}10^{13}$ photon/($cm^2$ s)  &     &  \\
 \hline
 Crab (Milagro)& 83.6 & 22.0 & 3.1 & $\infty$ & 91.1 &  229.9 & \cite{milagrocrabspectrum} \\
 Crab (Milagro)&      &      & 2.6 & 31 & 79.2 &  163.1 & \cite{milagrocrabspectrum} \\
 Crab (Reference)&      &      & 2.63 & $\infty$ & 68.5 &  147.8 & \cite{hesscrab} \\
 MGRO J2019+37 & 206.1 & 18.4 & 2.8 & $\infty$ & 40.5 &  66.8 & \cite{milagrocygnusspectrum} \\
 MGRO J1908+06 & 287.1 & 6.18 & 2.1 & $\infty$ & 17.6 & 36.6 & \cite{hess1908} \\
 MGRO J2031+41 & 308.0 & 41.6 & 3.2 & $\infty$ & 33.8 & 53.0&  \cite{milagrocygnusspectrum} \\
 MRK 421 (very low)& 166.1 & 38.2 &2.29  & 1.59 & 18.4 &  31.5 & \cite{mrk421variability} \\
 MRK 421 (mid)&  &  &2.28  & 4.36 & 131.5 &  178.0 & \cite{mrk421variability} \\
 MRK 421 (very high)&  &  &1.87  & 2.74 & 462.9 &  567.9 & \cite{mrk421variability}\\
 M87 & 187.7 & 12.2   & 2.31 & $\infty$ & 2.3 &  4.7 &  \cite{veritasm87} \\
 IC443 (hard) & 94.2 & 22.5 & 2.61  & $\infty$ & 2.6 &  5.6 & \cite{ic443veritas} \\
 IC443 (measured)   &      &      & 2.99  & $\infty$ & 1.1 &  2.5& \cite{ic443veritas} \\
 IC443 (soft)&      &      & 3.37  & $\infty$ & 0.4 &  1.3& \cite{ic443veritas}  \\
  \end{tabular}
  \caption{Table of a selection of known TeV sources (not complete) shown with the
    expected measurement significance after one year of HAWC data, assuming
    the sources are point sources, under
    specific hypotheses for the energy spectrum.
  }
  \label{potentialsources}
\end{sidewaystable}

\section{Discussion}

The HAWC instrument is designed to study particle acceleration in
Galactic and extra-Galactic sources as well as the propagation of
high-energy particles through the Galaxy and the extra-Galactic background
light (EBL).
Table \ref{potentialsources} illustrates the scientific potential of HAWC
by projecting a number of estimated HAWC measurements on a selection of
known TeV sources.

Pulsar Wind Nebulae (PWN) are the most common Galactic source of TeV
gamma rays \cite{pwn1}\cite{pwn2}.
A central rotating neutron star powers an electromagnetically-driven
flow of energetic electrons:
the pulsar wind. These electrons can radiate in the surround material
and can be further accelerated in shocks.

In a surprising twist to conventional thinking on PWN, recent flares
of the Crab Nebula have been observed at energies between 100 MeV and 100 GeV
by AGILE \citep{agilecrabflare}
and the Fermi-LAT
\citep{fermiflare}.
Above 1 TeV, the ARGO-YBJ collaboration has
announced possible evidence for variability \citep{argoflare} though
variability has not been yet observed by IACT experiments.
The Fermi data show strong transient flares in the synchrotron
emission from the
Crab Nebula, peaking at 50 times the quiescent state, over several days.

To account for the rapid flaring,
it has been suggested that the flares may be due to acceleration
by magnetic reconnection rather than shock acceleration 
(e.g. \cite{crabflarereconnection}).
Currently, we do not know if the flares
are confined to the synchrotron emission or whether they extend to the
higher-energy inverse-Compton emission. HAWC will be able to detect
an order-of-magnitude flare of the Crab in less than 10
minutes and will identify or
constrain the inverse-Compton emission from these flares. Furthermore,
it is currently not known whether other PWN flare like the Crab.
HAWC will continuously monitor other PWN for flares.

Supernova Remnants (SNR) \cite{snr}
have been observed to emit multi-TeV photons.
SNRs are the leading candidates for cosmic ray acceleration in the
Galaxy, and TeV emission can provide indirect evidence of hadronic
acceleration inside these objects.
Compelling evidence of cosmic ray acceleration
comes from the association of TeV photons from SNR interacting with
nearby molecular clouds, which would be inconsistent with TeV emission from inverse Compton
scattering.
As a specific example, the TeV emission from the SNR IC~443
measured by VERITAS\cite{ic443veritas}
has been interpreted
as the interaction of cosmic rays inside a molecular cloud.
This interpretation was strengthened by the observation of the 
characteristic neutral pion decay spectrum by Fermi\cite{fermipion}.
The measured spectral index of IC~443 above 1 TeV is relatively
uncertain but it will be detectable with HAWC after only a year of data-taking
if the spectrum is as hard as the VERITAS error bars allow. 
Particularly
intriguing for IC~443 is the Milagro 3$\rm{\sigma}$ excess at 35 TeV
\cite{milagrobsl},
which, if confirmed, is 10 times larger than anticipated
from extrapolating the central values of the
VERITAS measurement. Given the uncertainty
in the spectral index measured by VERITAS and the Milagro point, it may
be that the spectrum
from IC~443 is in fact harder than measured or that a second hard component is
measured by Milagro. HAWC will resolve this question for IC~443 and other SNR.

HAWC will study the TeV emission from Active Galactic Nuclei (AGN).
AGN emission can be extremely variable with flares up to ten times the
quiescent flux or more
(e.g. \cite{mrk421variability}).
The large distances to many AGN mean that HAWC data can be used to
study the far-infrared component of the EBL \citep{ebl1}, secondary
gamma-ray production between cosmic-ray sources and Earth \citep{ebl2}, and
even exotic modes of inter-galactic particle transport
such as axions\cite{axionagn}.
Furthermore, HAWC will provide unparalleled 
sensitivity to TeV AGN flares across the
entire overhead sky. This is particularly interesting given the occurrence
of so-called ``orphan'' flares in which the inverse-Compton 
emission from a blazar will 
flare without an accompanying flare in the 
synchrotron emission \citep{orphan1es1959}.
It has been suggested that these flares, defying explanation by the 
simplest Synchrotron Self-Compton models, may indicate the presence of
hadronic acceleration in the blazar \citep{orphantheory}\citep{orphantheory2}.
With IACTs blind to most of the overhead sky, and with no corresponding 
synchrotron
flare to trigger a search, an instrument like HAWC is needed
to observe these flares. Identified orphan flares can be used to trigger
searches in neutrino telescopes such as IceCube \citep{icecube}.
Detected contemporaneous neutrino emission would provide concrete evidence
of hadron acceleration in blazars.

HAWC data will also be used to study physics beyond the Standard Model.
For example, searches for gamma-ray emission from high-mass, low-luminosity
satellites of the Milky Way are a useful channel for detecting
dark matter particles
with masses over
1 TeV\cite{segue1fermi}\cite{segue1veritas}\cite{segue1magic}. If dark
matter interactions
are occurring in such objects, these processes will
likely result in the production of secondary gamma rays.
The production
of gamma rays from cosmic-ray acceleration and interaction in these
objects is expected to be very low, so an observation of gamma rays
may indicate the presence of dark matter interactions. Typical photon energy
spectra from the annihilation or decay of dark matter of mass $m_{\chi}$ are
concentrated at photon energies about a factor of 10 lower than
$m_{\chi}$. The superior sensitivity of HAWC above a photon energy of 1 TeV
implies
that HAWC will provide competitive limits on dark matter annihilation
and decay
for $m_{\chi}$ above $\sim$10 TeV, depending on the annihilation or decay channel.

\section{Conclusion}

Multi-TeV gamma rays are a probe of particle acceleration in astrophysical
sources. These sources can be transient, flaring by an order of magnitude
or more in a matter of hours, with spectra
already measured to 20 TeV or more with current instruments. 
The High Altitude Water Cherenkov
observatory is sensitive to photons from 100 GeV to 100 TeV, with a peak
sensitivity at the 10-20 TeV, where existing IACT spectra end.
HAWC will observe the entire overhead sky for transient emission and
extend and constrain spectra up to 100 TeV. HAWC will survey
the sky with a point-source 
sensitivity of
$5 \times 10^{-13}$ $\rm{cm}^{-2} \rm{sec}^{-1}$ above 2 TeV
in a year of data-taking across 5 sr
(40\%) of the sky.
When completed, HAWC will be the most sensitive
gamma-ray detector above 10 TeV
and will maintain unprecedented wide-field sensitivity to gamma rays
above 100 GeV.

\section*{Acknowledgments}

We gratefully acknowledge Scott DeLay
his dedicated efforts in the construction and
maintenance of the HAWC experiment.
This work has been supported by: the National Science Foundation, 
the US Department of Energy Office of High-Energy Physics,
the LDRD program of Los Alamos National Laboratory, 
Consejo Nacional de Ciencia y Tecnologia 
(grants 55155, 103520, 105033,
105666, 122331 and 132197), 
Red de Fisica de Altas Energias,
DGAPA-UNAM (grants IN105211, IN112910 and IN121309,
IN115409), VIEP-BUAP (grant 161-EXC-2011), the University
of Wisconsin Alumni Research Foundation, and the Institute of 
Geophysics and Planetary Physics at Los Alamos National Lab.
\bibliography{bibliography}

\end{document}

%% file: hawc-authors-elsevier-initials.tex
\author[MSU]{A. U. Abeysekara}
\author[IF-UNAM]{R. Alfaro}
\author[UNACH]{C. Alvarez}
\author[UMSNH]{J. D. {\'A}lvarez}
\author[UNACH]{R. Arceo}
\author[UMSNH]{J. C. Arteaga-Vel{\'a}zquez}
\author[MTU]{H. A. Ayala Solares}
\author[University of Utah]{A. S. Barber}
\author[UMD]{B. M. Baughman \corref{cor1}}
\author[UPP]{N. Bautista-Elivar}
% ...

\address[MSU]{Department of Physics and Astronomy, Michigan State University, East Lansing, MI, USA}
\address[IF-UNAM]{Instituto de F{\'\i}sica, Universidad Nacional Aut{\'o}noma de M{\'e}xico, Mexico D.F., Mexico}
\address[UNACH]{CEFyMAP, Universidad Aut{\'o}noma de Chiapas, Tuxtla Guti{\'e}rrez, Chiapas, Mexico}
\address[UMSNH]{Universidad Michoacana de San Nicol{\'a}s de Hidalgo, Morelia, Mexico}
\address[MTU]{Department of Physics, Michigan Technological University, Houghton, MI, USA}
\address[University of Utah]{Department of Physics and Astronomy, University of Utah, Salt Lake City, UT, USA}
\address[UMD]{Department of Physics, University of Maryland, College Park, MD, USA}
\address[UPP]{Universidad Politecnica de Pachuca, Pachuca, Hgo, Mexico}
% ...
\address[IGeof-UNAM]{Instituto de Geof{\'\i}sica, Universidad Nacional Aut{\'o}noma de M{\'e}xico, Mexico D.F., Mexico}
% ...
\address[IA-UNAM]{Instituto de Astronom{\'\i}a, Universidad Nacional Aut{\'o}noma de M{\'e}xico, Mexico D.F., Mexico}
% ...
\address[ICN-UNAM]{Instituto de Ciencias Nucleares, Universidad Nacional Aut{\'o}noma de M{\'e}xico, Mexico D.F., Mexico}
% ...

\cortext[cor2]{jpretz@lanl.gov}
\cortext[cor1]{bbaugh@umdgrb.umd.edu}

%% file: bins.tex
1 & 39-59 & 1.0-7.0 & 1.30 &  3.1 & $\rm{5.7x10^{4}}$ & $\rm{2.8x10^{-3}}$ &  2.5 & 0.46 \\
2 & 60-69 & 1.0-7.0 & 1.00 &  5.2 & $\rm{1.4x10^{4}}$ & $\rm{9.6x10^{-3}}$ &  2.6 & 0.47 \\
3 & 70-90 & 1.0-7.0 & 0.88 &  5.3 & $\rm{1.8x10^{4}}$ & $\rm{1.3x10^{-2}}$ &  2.7 & 0.44 \\
4 & 91-147 & 1.0-7.0 & 0.68 &  8.1 & $\rm{1.8x10^{4}}$ & $\rm{4.4x10^{-2}}$ &  2.9 & 0.40 \\
5 & 148-231 & 1.0-7.0 & 0.50 & 11.7 & $\rm{7.9x10^{3}}$ & $\rm{1.7x10^{-1}}$ &  3.0 & 0.35 \\
6 & 232-349 & 1.0-7.0 & 0.36 & 13.4 & $\rm{3.7x10^{3}}$ & $\rm{4.9x10^{-1}}$ &  3.2 & 0.32 \\
7 & 350-495 & 1.0-7.0 & 0.30 & 17.2 & $\rm{1.4x10^{3}}$ & $\rm{2.0}$ &  3.5 & 0.28 \\
8 & 496-655 & 1.0-7.0 & 0.22 & 17.7 & $\rm{6.0x10^{2}}$ & $\rm{4.7}$ &  3.7 & 0.24 \\
9 & 656-789 & 1.0-7.0 & 0.20 & 17.1 & $\rm{2.4x10^{2}}$ & $\rm{9.8}$ &  3.8 & 0.21 \\
10 & 790-1200 & 4.0-4.5 & 0.16 & 14.4 & $\rm{1.4x10^{2}}$ & $\rm{1.6x10^{1}}$ &  4.0 & 0.18 \\
11 & 790-1200 & 4.5-4.9 & 0.14 & 11.5 & $\rm{1.2x10^{2}}$ & $\rm{1.2x10^{1}}$ &  4.2 & 0.18 \\
12 & 790-1200 & 4.9-5.3 & 0.12 &  7.2 & $\rm{2.5x10^{1}}$ & $\rm{1.2x10^{1}}$ &  4.6 & 0.07 \\
13 & 790-1200 & 5.3-5.7 & 0.12 &  1.9 & $\rm{3.4}$ & $\rm{1.8x10^{-1}}$ &  5.1 & 0.13 \\
14 & 790-1200 & 5.7-6.4 & 0.08 &  0.9 & $\rm{3.2x10^{-1}}$ & $\rm{3.6x10^{-1}}$ &  5.5 & 0.10 \\